\documentclass[letterpaper,11pt]{article} 
\usepackage{graphics,graphicx}
\usepackage{longtable}
\usepackage{amsmath,verbatim,amssymb,epsfig}
\usepackage{wrapfig}
\usepackage{amsmath,verbatim,amssymb,epsfig,color,lscape}
\usepackage{natbib}
\usepackage{multirow,multicol}
\usepackage{authblk}
\usepackage[pagestyles]{titlesec}
\usepackage[normalem]{ulem}
\usepackage{bm}
\usepackage{url}
\newcommand{\bX}{\bm{X}}
\newcommand{\bY}{\bm{Y}}
\newcommand{\bU}{\bm{U}}
\newcommand{\bD}{\bm{D}}
\newcommand{\hide}[1]{}

\newcommand{\bi}{\begin{itemize}}
\newcommand{\ei}{\end{itemize}}
\newcommand{\mbP}{\mathbb{P}}
\newcommand{\mR}{\mathbb{R}}

\makeatletter
\newcommand{\s}{\vspace{.25cm}}

\newcommand{\msim}{\mathop{\rm \sim}}

\usepackage[outerbars,color]{changebar}
\ifx\pdfoutput\undefined
\else\ifnum\pdfoutput>0
  \usepackage{pdfcolmk}
\fi\fi
\cbcolor{black}

\usepackage[margin=1.0in]{geometry}
\newcommand{\ind}{\msim\limits^{\mbox{\tiny ind}}}
\newcommand{\iid}{\msim\limits^{\mbox{\tiny iid}}}

\newcommand{\be}{\begin{equation}\begin{array}{lllllllllllll}\displaystyle}
\newcommand{\beno}{\begin{equation}\begin{array}{lllllllllllll}\nonumber\displaystyle}

\newcommand{\ee}{\end{array}\end{equation}}

\newfont{\rmm}{cmr10 at 11pt}
\rmm

\begin{document}

\title{Multilevel Network Item Response Modeling for Discovering Differences Between Innovation and Regular School Systems in Korea}
\author[1,2]{Ick Hoon Jin}
\author[3]{Minjeong Jeon}
\author[4]{Michael Schweinberger}
\author[5]{Jonghyun Yun}
\author[6]{Lizhen Lin}
\affil[1]{Department of Applied Statistics, Yonsei University}
\affil[2]{Department of Statistics and Data Science, Yonsei University}
\affil[3]{School of Education and Information Studies, University of California, Los Angeles}
\affil[4]{Department of Statistics, University of Missouri, Columbia}
\affil[5]{Institute of Statistical Data Intelligence}
\affil[6]{Department of Applied and Computational Mathematics and Statistics, University of Notre Dame}
\date{}
\maketitle

\begin{abstract}
The innovation school system in South Korea has been developed in response to the traditional high-pressure school system in South Korea,
with a view to cultivating a bottom-up and student-centered educational culture.
Despite its ambitious goals,
questions have been raised about the success of the innovation school system.
Leveraging data from the Gyeonggi Education Panel Study (GEPS) along with advances in the statistical analysis of network data and educational data,
we compare the two school systems in more depth.
We find that some schools are indeed different from others,
and those differences are not detected by conventional multilevel models.
Having said that,
we do not find much evidence that the innovation school system differs from the regular school system in terms of self-reported mental well-being,
although we do detect differences among some schools that appear to be unrelated to the school system.
\end{abstract}

\noindent%

\section{Introduction}
\label{sec:intro}

The South Korean public K-12 system has long been criticized for its competitive environment that is believed to be detrimental to the mental and physical health, autonomy, creativity, and democratic conscience of students.
As a remedy, 
the Gyeonggi Province Office of Education introduced an innovation school program in 2009,
with a view to cultivating a bottom-up and student-centered culture \citep{Gyeonggi:2012}.
The innovation school program was designed to (1) provide schools and teachers with more autonomy in the choice of teaching materials and methods;
(2) foster creative and self-directed learning;
and (3) encourage honest communication and mutual respect.
Since its launch in 2009,
the innovation school program has been expanded by the South Korean Ministry of Education \citep{Gu:2013, Lee:2012}.

However,
the innovation school program has received mixed reviews \citep[e.g.,][]{Bae:14, Baek:14, Kim:11}.
Critics have pointed out that the innovation school system fails to improve the academic performance of students \citep[e.g.,][]{Kim:11, Baek:14}.  %\textcolor{red}{(reference)}.
Advocates have claimed that the innovation school system should be evaluated based on non-cognitive outcomes rather than cognitive outcomes, 
because the system aims to stimulate non-cognitive skills, such as creativity, ethics, and autonomy,
in addition to reducing stress and improving mental well-being \citep[e.g.,][]{Nah:13, Min:17}. %\textcolor{red}{(reference)}.
Researchers have attempted to measure non-cognitive outcomes of the innovation school program,
with varied results \citep{Jang:14, Kim:14, Kim:16, Cho:16}.
For instance, 
\citet{Nah:13} reported that the innovation school program has had a positive effect on psychological attributes of students,
such as self-esteem, academic efficacy, stress, and depression. 
\citet{Min:17} showed that innovation school students expressed greater satisfaction with school than regular school students.
On the other hand, 
\citet{Sung:14} reported that there are few differences in the academic stress level and class attitudes between students of the two school systems.
An issue of many of these analyses is that these analyses rely on traditional comparisons between the two school programs,
using analysis of variance and $t$-tests \citep{Min:17}.

Leveraging data from the Gyeonggi Education Panel Study (GEPS) along with advances in the statistical analysis of network data and educational data, 
we compare the two school systems in more depth.
The GEPS data set is a large-scale 
% representative panel 
survey of K-12 students in Gyeonggi province and offers a unique opportunity to evaluate the innovation school system, 
for at least two reasons: 
First, 
Gyeonggi province -- the second-largest province of South Korea -- was the first to implement the innovation school system and is hence a natural starting point for exploring differences between the two school systems.
Second, 
the GEPS data includes both regular and innovation schools,
enabling a comparison of the innovation school system with the regular school system.
We analyze these data by combining recent advances in network-based approaches to educational data \citep{Jin:18} with a simple approach to multilevel data,
that is,
data collected from multiple schools that have implemented the regular or the innovation school system.

The remainder of our paper is structured of follows.
In Section 2, 
we describe the GEPS data.
In Section \ref{sec:DLSJM}, 
we introduce the modeling framework and compare it to existing approaches.
In Section 4, 
we present an application of the proposed approach to the GEPS data.

\section{GEPS data} \label{sec:description}

\paragraph*{Data description}

The 2009 GEPS data were based on a total of 3,918 tenth-graders from 62 high schools in the Gyeonggi province of South Korea (https://www.gie.re.kr/eng/content/C0012-04.do).
The students were sampled from 16 innovation and 46 regular schools to reflect the student and school population in the Gyeonggi province.  
For data analysis, we chose the third wave of the GEPS data that include third-year general high school (twelth-grade) students who had experienced three years of the innovation or regular school program. This choice was intended such that we could compare the outcomes of the students who had been taught under the regular and innovation school program (which covers entire high school curriculum). As the result, 16 innovation schools (with 904 students) and 46 regular schools (with 3,014 students) were included in the final dataset. For the sake of simplicity, students who transferred to different schools between 2009 and 2012 were excluded from the data analysis. %[is this true?]

\begin{figure}[htbp]
\centering
\includegraphics[width=0.6\textwidth]{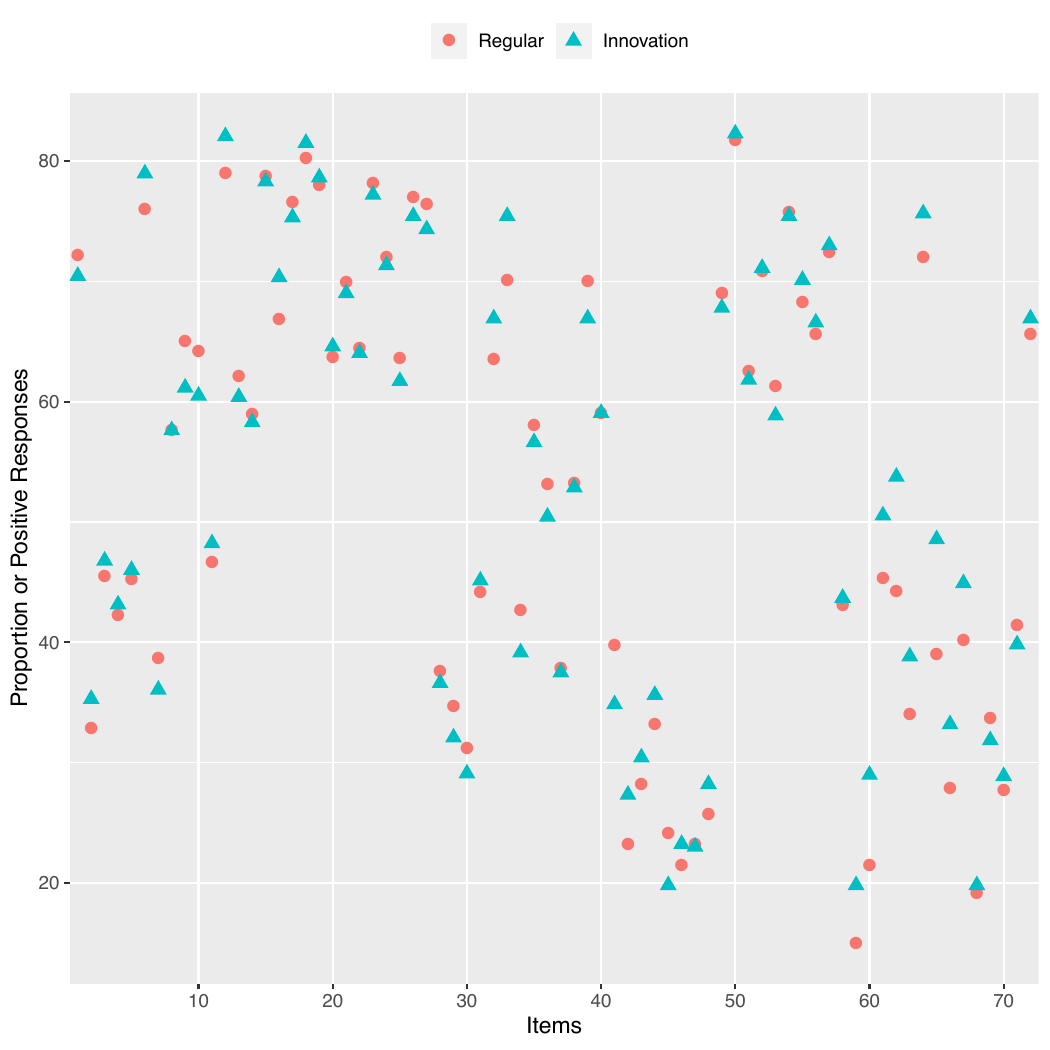}
\caption{\label{fig:item_score}
The proportion of positive responses for each of the 72 test items across all students of regular and innovation schools.
The labels of the horizontal and vertical axes indicate item numbers.}
\end{figure}

To evaluate students' non-cognitive outcomes, we selected 10 psychological/attitude scales that include Mental Ill-being (Item 1 - 6), Sense of Citizenship (Item 7 - 22), Self-Efficacy (Item 23 - 30, 71), Disbelief in Growth (Item 31 - 33), Self-Driven Learning (Item 34 - 37), Self-Understanding (Item 38 - 41), Test Stress (Item 42 - 48), Relationship with Friends (Item 49 - 54), Self-Esteem (Item 55 - 58, 72), and Academic Stress (Item 59 - 70). These are not established scales; hence, the trait measurements from these scales may be theoretically nonequivalent to the traits implied by the scale names. Each scale includes three to thirteen items that are measured with a five-point agreement-based Likert scale. 
All items are presented in the supplementary materials (Section A).\footnote{The original items are formulated in Korean. The English translations of items presented in the remainder of the paper are unofficial translations by MJ and IHJ.} 
Some of the items were negatively worded, and in that case we reverse-coded the responses to facilitate the interpretation of results (Items 1--6, 20--22, 31--33, 42--48, and 58--70).  
The responses from all individual items were  then   dichotomized such that 1 represents positive responses (``agree'') while 0 represents negative responses (``disagree'') for the data analysis. 

Figure \ref{fig:item_score} displays the proportion of positive responses for each of the 72 test items across all students within the regular and innovation school systems. The figure shows that the degree and direction of the differences in the proportion scores widely varied across individual items. 
This suggests that univariate analysis using a summary score (e.g., mean, sum) can be misleading as such variation may be lost during the summarizing process. 
In addition, as aforementioned, the 72 items are from 10 different scales that measure related, but different psychological or attitudinal attributes of the students. All items are likely to be correlated with each other in varying degrees, 
while such correlation structures may  differ between the two school systems. 
We propose a method that does not assume that all items have the same threshold levels and that the items are independent of each other. 

\paragraph*{Scope and goal of the data analysis}

To analyze the GEPS data,
we use all 72 items without distinguishing between scales,
which is reasonable as the scales are not well-established scales but working scales.
Our proposed approach explores similarities and dissimilarities of items and identifies clusters of items based on similarities and dissimilarities of the items. 
It is worth noting that we that do not discuss the sources of the differences between the school systems nor provide causal explanations.
Causal inference based on observational data is non-trivial and is beyond the scope of our paper.
Our data analysis is an exploratory approach aiming to identify  differences between the two school systems,
without attempting to make causal statements.

\section{Multilevel network item response model}\label{sec:DLSJM}

We introduce a multilevel network item response model with a view to analyzing the GEPS data,
building on latent space models.
To prepare the ground for the proposed modeling framework,
we first provide a concise introduction to latent space models in Section \ref{lsm} and then introduce the proposed modeling framework in Section \ref{prop1}.
Throughout,
we denote the set of real numbers by $\mR$ and the set of positive real numbers by $\mR_+$.
The Euclidean distance between two vectors $\bm{a}$ and $\bm{b}$ in $\mR^q$ is denoted by 
\beno
d(\bm{a},\, \bm{b}) 
&=& \sqrt{\displaystyle\sum_{i=1}^q (a_i - b_j)^2},
&\; \bm{a} \in \mR^q,
& \bm{b} \in \mR^q,
& q \geq 1.
\ee

\subsection{Latent space models}
\label{lsm}

The idea of embedding data into Euclidean space has a long history in educational statistics and related areas,
as demonstrated by multidimensional scaling \citep[e.g.,][]{To52,Tw65,Oh:2001,Oh:2007}.
In recent decades,
the idea of embedding data into Euclidean space has been adapted to network data represented by a graph with connections among nodes \citep{Hoff:2002}.
Known as latent space models,
the basic idea is that nodes $i$ and $j$ have positions $\bm{z}_i$ and $\bm{z}_j$ in $\mR^q$ and the probability of a connection $Y_{i,j} \in \{0, 1\}$ between $i$ and $j$ depends on the distance $d(\bm{z}_i,\, \bm{z}_j)$ between them:
\beno
Y_{i,j} \mid \beta,\, d(\bm{z}_i,\, \bm{z}_j)
&\ind& \mbox{Bernoulli}(\mbP(Y_{i,j} = 1 \mid \beta,\, d(\bm{z}_i,\, \bm{z}_j))),
\ee
where
\beno
\log \displaystyle\frac{\mbP(Y_{i,j} = 1 \mid \beta,\, d(\bm{z}_i,\, \bm{z}_j))}{1-\mbP(Y_{i,j} = 1 \mid \beta,\, d(\bm{z}_i,\, \bm{z}_j))}
&=& \beta - d(\bm{z}_i,\, \bm{z}_j).
\ee
In other words,
the log odds of a connection between nodes $i$ and $j$ is $\beta \in \mR$ when the distance $d(\bm{z}_i,\, \bm{z}_j)$ separating $i$ and $j$ in $\mR^q$ is $0$,
and is otherwise $\beta - d(\bm{z}_i,\, \bm{z}_j)$.
Thus,
the greater the distance separating nodes $i$ and $j$ in $\mR^q$ is,
the lower is the log odds of a connection between them.
It is a convention to choose the dimension $q$ of $\mR^q$ to be 2 or 3,
to facilitate 2- or 3-dimensional representations of nodes in $\mR^2$ or $\mR^3$,
respectively.
If the random graph represents a social network,
the latent space may be interpreted as an unobserved social space underlying the observed network of interactions.
More background on latent space models can be found in,
e.g.,
\cite{Hoff:2002,ScSn03,Handcock:2007, Krivitsky:2009, Raftery:2012, latentspace, Sweet:13, STMu13,TaSuPr13,SuTaPr14,sewell2015analysis,sewell2015latent,sewell2016latent,FoHo15,Rastelli:2015,Gollini:2016,SmAsCa19}; 
and \cite{MaMaYu20}.

\subsection{Proposed modeling framework}
\label{prop1}

We first review models of individual schools and then discuss multilevel extensions.

To introduce models of individual schools,
we follow \cite{Jin:18} and introduce two sets of networks that represent similarities of students and items in a school of interest,
and then adapt the approach of \cite{Jin:18} to multilevel data -- that is,
data from multiple schools.
Consider $M$ schools $1, \dots, M$,
with responses $X_{m,i,k} \in \{0, 1\}$ to items $i \in \{1, \dots, p\}$ ($p \geq 1$) by students $k \in \{1, \dots, n_m\}$ ($n_m \geq 1$) in school $m \in \{1, \dots, M\}$ ($M \geq 1$).
We consider two sets of networks,
depending on whether embedding students or items in $\mR^q$ is of primary interest:
\begin{itemize}
\item {\bf If embedding students in $\mR^q$ is of primary interest,
construct networks $\bY_{m,i}$ as follows} ($i = 1, \dots, p$,\, $m = 1, \dots, M$):
\[
\bY_{m,i}\; =\; (Y_{m,i,k,l})\; =\; (X_{m,i,k}\, X_{m,i,l}),
\]
where $Y_{m,i,k,l} = 1$ indicates that students $k$ and $l$ in school $m$ both agree with item $i$.
\item {\bf If embedding items in $\mR^q$ is of primary interest,
construct networks $\bU_{m,k}$ as follows} ($k = 1, \dots, n_m$,\, $m = 1, \dots, M$):
\[
\bU_{m,k}\; =\; (U_{m,k,i,j})\; =\; (X_{m,i,k}\, X_{m,j,k}),
\]
where $U_{m,k,i,j} = 1$ indicates that student $k$ in school $m$ agrees with both items $i$ and $j$.
\end{itemize}
These networks are all undirected,
in the sense that connections in these networks do not have directions.
It is worth noting that there is more than one approach to constructing networks.
The construction of networks described above makes some sense, 
in that students are embedded based on shared agreements with items,
whereas items are embedded based on shared agreements by individual students.
That being said,
there are other possible constructions of networks:
e.g.,
to embed items,
one could construct networks $\bU_{m,k}^\star$ by definining
\beno
U_{m,k,i,j}^\star 
&=& X_{m,i,k}\, X_{m,j,k} + (1 - X_{m,i,k})\, (1 - X_{m,j,k}),
\ee
in which case items $i$ and $j$ would be considered similar if student $k$ in school $m$ either agreed with both items or disagreed with both items.
Indeed,
one of the advantages of the proposed modeling framework is its flexibility in the construction of networks,
which allows researchers to embed either items or students based on similarity measures of interest.

\paragraph*{Embedding students based on latent space models of $\bY$'s}

Students can be embedded in $\mR^q$ by assuming that students $k$ and $l$ have positions $\bm{z}_{m,k}$ and $\bm{z}_{m,l}$ in $\mR^q$ and that,
conditional on the distance $d(\bm{z}_{m,k},\, \bm{z}_{m,l})$ separating students $k$ and $l$ in $\mR^q$,
the random variables $Y_{m,i,k,l}$ are independent Bernoulli random variables,
with log odds
\be
\log\frac{\mbP(Y_{m,i,k,l} = 1 \mid \beta_{m,i},\, d(\bm{z}_{m,k},\, \bm{z}_{m,l}))}{1 - \mbP(Y_{m,i,k,l} = 1 \mid \beta_{m,i},\, d(\bm{z}_{m,k},\, \bm{z}_{m,l}))} 
&=& \beta_{m,i} - d(\bm{z}_{m,k},\, \bm{z}_{m,l}).
\label{eq:model1}
\ee
The parameter $\beta_{m,i} \in \mR$ can be interpreted as the log odds that both students $k$ and $l$ in school $m$ agree with item $i$,
provided that $k$ and $l$ are separated by distance $d(\bm{z}_{m,k},\, \bm{z}_{m,l}) = 0$.
Otherwise,
the log odds is reduced by the distance between students $k$ and $l$.

\paragraph*{Embedding items based on latent space models of $\bU$'s}

Along the same lines,
items can be embedded in $\mR^q$ by assuming that items $i$ and $j$ have positions $\bm{w}_{m,i}$ and $\bm{w}_{m,j}$ in $\mR^q$ and that,
conditional on the distance $d(\bm{w}_{m,i},\, \bm{w}_{m,j})$ separating items $i$ and $j$ in $\mR^q$,
the random variables $U_{m,k,i,j}$ are independent Bernoulli random variables,
with log odds
\be
\log\frac{\mbP(U_{m,k,i,j} = 1 \mid \theta_{m,k},\, d(\bm{w}_{m,i},\, \bm{w}_{m,j}))}{1 - \mbP(U_{m,k,i,j} = 1 \mid \theta_{m,k},\, d(\bm{w}_{m,i},\, \bm{w}_{m,j}))} 
&=& \theta_{m,k} - d(\bm{w}_{m,i},\, \bm{w}_{m,j}).
\label{eq:model2}
\ee
The parameter $\theta_{m,k} \in \mR$ corresponding to student $k$ in school $m$ is the log odds that student $k$ agrees with both items $i$ and $j$ when items $i$ and $j$ are separated by distance $d(\bm{w}_{m,i},\, \bm{w}_{m,j}) = 0$.
Otherwise,
the log odds is reduced by the distance between items $i$ and $j$.

\paragraph*{Joint probability model of $\bY$'s and $\bU$'s}

It is worth noting that the latent space models described above are proper probability models of the $\bY$'s and $\bU$'s,
but it is not straightforward to construct a joint probability model of the $\bY$'s and $\bU$'s:
e.g.,
it is not possible to construct a joint probability model of the $\bY$'s and $\bU$'s by assuming that the $\bY$'s and $\bU$'s are conditionally independent given the positions of students and items in $\mR^q$,
because the $\bY$'s and $\bU$'s are deterministic functions of $\bX$.
That said,
we are free to use the latent space models described above to embed either students or items in $\mR^q$.
If it is desired to embed both students and items in $\mR^q$ to shed light on how students interact with items,
the recent approach of \citet{jeon:20} can be used.

\paragraph*{Identifiability issues}

The described latent space models have the same identifiability issues as other latent space models, 
rooted in the invariance of Euclidean distances to translation, reflection, and rotation of the positions of students and items in $\mR^q$. 
As a consequence, 
the log odds \eqref{eq:model1} and \eqref{eq:model2} are invariant to these transformations,
as is the resulting likelihood function.
We follow the conventional approach to addressing these identifiability issues by basing statistical inference on equivalence classes of positions,
using Procrustes matching \citep{Hoff:2002}. 

\paragraph*{Selecting the dimension $q$ of $\mR^q$}

In principle, 
the dimension $q$ of $\mR^q$ can be determined by model selection criteria,
such as Bayesian information criteria \citep{Handcock:2007},
although the theoretical properties of Bayesian information criteria in applications to latent space models are not well-understood.
We therefore follow convention and choose $q=2$ \citep{Hoff:2002},
which has advantages in terms of model parsimony and visualization. 

\paragraph*{Multilevel extensions}

To discuss a simple multilevel extension of the latent space models introduced above,
suppose that it is desired to embed items in $\mR^q$ and combine the results across the $M$ schools.
To combine the results across the $M$ schools and embed the $M$ schools in $\mR^q$ along the way,
construct a $M \times M$ matrix $\bm{\Delta}$ of dissimilarities $\bm{\Delta}_{a,b} = \sqrt{\sum_{i<j}^p (d_{a,i,j} - d_{b,i,j})^2}$ between schools $a$ and $b$,
where $d_{m,i,j}$ refers to the distance $d(\bm{w}_{m,i},\, \bm{w}_{m,j})$ between items $i$ and $j$ in school $m \in \{1, \dots, M\}$.
We can then embed the $M$ schools in $\mR^q$ based on how dissimilar the schools are in terms of item responses by students,
by applying Kruskal's multidimensional scaling procedure to the dissimilarity matrix $\bm{\Delta}$ \citep[][]{Kr64,Cox:01}.
The resulting procedure has the advantage that it requires nothing more than postprocessing the posterior samples of the $M$ schools and is therefore a simple alternative to random effects approaches to multilevel models,
and provides an embedding of schools in $\mR^q$.

\subsection{Priors}
 
We use the following priors.
The hyperparameters of these priors are specified in Section 4.

\paragraph*{Item parameters}

Let $\boldsymbol\beta_m = (\beta_{m,1}, \cdots, \beta_{m,p}) \in \mR^p$ denote the item parameter vector specific to school $m$ ($m = 1, \cdots, M$).
We assume that the $M$ item parameter vectors $\boldsymbol\beta_1, \cdots, \boldsymbol\beta_M$ are independent,
with prior
\beno
\beta_{m,i} \mid \sigma_{\beta}^2 &\iid& N(0,\, \sigma_\beta^2),
\ee
where $\sigma_{\beta}^2 \in \mR_+$ is a specified hyperparameter.

\hide{

\paragraph*{Item parameters}

Let $\boldsymbol\beta_m = (\beta_{m,1}, \cdots, \beta_{m,p}) \in \mR^p$ denote the item parameter vector specific to school $m$ ($m = 1, \cdots, M$).
We assume that the $M$ item parameter vectors $\boldsymbol\beta_1, \cdots, \boldsymbol\beta_M$ are independent and have the following prior:
\beno
\beta_{m,i} \mid \gamma_i,\, \sigma_{i}^2 &\iid& N(\gamma_i,\, \sigma_{i}^2)\s
\\
\gamma_{i} \mid \sigma_{\gamma}^2 &\iid& N(0,\, \sigma_{\gamma}^2)\s
\\
\sigma_i^2 \mid a_{\sigma},\, b_{\sigma} &\iid& \mbox{Inverse-Gamma}(a_{\sigma},\, b_{\sigma}),
\ee
where $\sigma_{\gamma}^2 \in \mR_+$, $a_{\sigma} \in \mR_+$, and $b_{\sigma} \in \mR_+$ are specified hyperparameters.
We write $\bm\gamma = (\gamma_1, \cdots, \gamma_p) \in \mR^p$ and $\bm\sigma^2 = (\sigma_1^2, \cdots, \sigma_p^2) \in \mR_+^p$.

}

\paragraph*{Student parameters}

Let $\boldsymbol\theta_m = (\theta_{m,1}, \cdots, \theta_{m,n_m}) \in \mR^{n_m}$ denote the student parameter vector specific to school $m$ ($m = 1, \cdots, M$).
We assume that the $M$ student parameter vectors $\boldsymbol\theta_1, \cdots, \boldsymbol\theta_M$ are independent,
with prior
\beno
\theta_{m,k} \mid \sigma_{\theta}^2 
&\iid & N(0,\, \sigma_{\theta}^2),
\ee
where $\sigma_{\theta}^2 \in \mR_+$ is a specified hyperparameter.

\paragraph*{Item distances}

Let $d_{m,i,j} \equiv d(\bm{w}_{m,i},\, \bm{w}_{m,j})$ be the distance between items $i$ and $j$ in school $m$,
and let $\bD_{m} = (d_{m,i,j})$ be the $p\times p$ matrix of distances between items in school $m$ ($m = 1, \dots, M$).
We allow variation across the $M$ schools by using the following prior:
\beno
d_{m,i,j} \mid \mu_{i,j},\, \tau_{i,j}^2 &\ind& \mbox{Lognormal} (\mu_{i,j},\, \tau_{i,j}^2),
\;\;\;\; i<j \in \{1, \dots, p\}\s
\\
\mu_{i,j} \mid \sigma_\mu^2 &\iid& N(0,\, \sigma_\mu^2)\s
\\
\tau_{i,j}^2 \mid a_\tau,\, b_\tau &\iid& \mbox{Inverse-Gamma}(a_\tau,\, b_\tau),
\ee
where $\sigma_\mu^2 \in \mR_+$, $a_\tau \in \mR_+$ and $b_\tau \in \mR_+$ are specified hyperparameters.
The collections of means $\mu_{i,j}$ and variances $\tau_{i,j}^2$ are denoted by $\bm\mu$ and $\bm\tau$,
respectively.
In addition to the single-group model described above,
we consider a multiple-group model,
which allows the means and variances to depend on the regular and innovation school system.
It is worth noting that the prior of the distances specified above is a convenience prior,
covering distances (which satisfy the triangle inequality) along with non-distances (which do not satisfy the triangle inequality).
While the prior is motivated by convenience,
it does cover the space of interest (the space of all distances) and the Bayesian algorithm described in Section 3.4 with postprocessing procedures (a) and (b) reports distances between items (satisfying the triangle inequality).
The use of convenience priors is not unprecedented in Bayesian statistics:
e.g.,
uniform priors may not reflect prior knowledge and may not even be proper priors,
but can give rise to proper posteriors and can be useful in applications \citep[e.g.,][]{improper.priors}.

\subsection{Bayesian inference}\label{sec:posterior}

We follow a Bayesian approach to estimating the multilevel network item response model and focus on embedding items in $\mR^q$ along with schools,
motivated by the application to the GEPS data.

\paragraph*{Posterior}

To embed items in $\mR^q$ along with schools,
compute the networks $\bU_{m,k} = (U_{m,k,i,j})$ by computing $U_{m,k,i,j} = X_{m,i,k}\, X_{m,j,k}$ and write $\bU_{m} = (\bU_{m,1}, \dots, \bU_{m,n_m})$ ($m = 1, \dots, M$).
%where $U_{m,k,i,j} = 1$ indicates that student $k$ in school $m$ agrees with both items $i$ and $j$.
Then the posterior of $\bm\theta_1, \dots, \bm\theta_M$ and $\bD_1, \dots, \bD_M$ given $\bU_1, \dots, \bU_M$ is proportional to
\beno
\pi(\bm\theta_1, \dots, \bm\theta_M,\, \bD_1, \dots, \bD_M \mid \bU_1, \dots, \bU_M) 
&\propto& \left[\displaystyle\prod_{m=1}^M\, \displaystyle\prod_{k=1}^{n_m}\, \displaystyle\prod_{i<j}^p\, \mbP(U_{m,k,i,j} = u_{m,k,i,j} \mid \theta_{m,k},\, d_{m,i,j})\right]\s
\\
&\times& \left[\displaystyle\prod_{m=1}^M\, \pi(\bm\theta_m)\; \pi(\bD_m \mid \bm\mu,\, \bm\tau)\right]\; \pi(\bm\mu)\; \pi(\bm\tau),
\ee
where $\bD_m = (d_{m,i,j})$ and $d_{m,i,j}$ refers to $d(\bm{w}_{m,i},\, \bm{w}_{m,j})$ ($m = 1, \dots, M$).\s

\vspace{-0.5cm}

\paragraph*{Markov chain Monte Carlo algorithm} 

A Markov chain Monte Carlo algorithm for sampling from the posterior distribution can then be constructed by combining the following Markov chain Monte Carlo steps, assuming all unknown quantities have been initialized.

{
\begin{enumerate}
\item[] {\bf Preprocessing:} 
Construct $\bU_1, \dots, \bU_M$.
\item[] {\bf Iterate}:
\item Sample the positions of items and the parameters of school $m$,
where $\bm{w}_{m,-i}$ refers to the positions of all items excluding $\bm{w}_{m,i}$ and $\bm\theta_{m,-k}$ refers to all parameters excluding $\theta_{m,k}$ ($m = 1, \dots, M$):
    \begin{enumerate}
    \item Sample $\bm{w}_{m,i} \mid \bU_m,\, \bm{w}_{m,-i},\, \bm\theta_m,\, \bm\mu,\, \bm\tau^2$ ($i = 1, \dots, p$,\, $m = 1, \dots, M$).
    \item Sample $\theta_{m,k} \mid \bU_m,\, \bm\theta_{m,-k},\, \bm{w}_{m,1},\, \dots,\, \bm{w}_{m,p}$ ($k = 1, \dots, n_m$, $m = 1, \dots, M$).
    \end{enumerate}
\item Sample the hyperparameters:
    \begin{enumerate}
    \item Sample $\mu_{i,j} \mid d(\bm{w}_{1,i},\, \bm{w}_{1,j}), \cdots, d(\bm{w}_{M,i},\, \bm{w}_{M,j}),\, \tau_{i,j}^2$.
    \item Sample $\tau_{i,j}^2 \mid d(\bm{w}_{1,i},\, \bm{w}_{1,j}), \cdots, d(\bm{w}_{M,i},\, \bm{w}_{M,j}),\, \mu_{i,j}$.
    \end{enumerate}
\item[] {\bf Postprocessing:} 
\begin{itemize}
\item[(a)] To solve the identifiability issue arising from the invariance of the likelihood function to reflection,
translation,
and rotation of the item positions,
use Procrustes matching. 
\item[(b)] To embed the $M$ schools in $\mR^q$,
construct a $M \times M$ matrix $\bm{\Delta}$ of dissimilarities\break
$\bm{\Delta}_{a,b} = \sqrt{\sum_{i<j}^p (d_{a,i,j} - d_{b,i,j})^2}$ between schools $a$ and $b$,
where $d_{m,i,j}$ refers to the distance $d(\bm{w}_{m,i},\, \bm{w}_{m,j})$ between items $i$ and $j$ in school $m \in \{1, \dots, M\}$ obtained in Step (a).
Then embed the $M$ schools in $\mR^q$ by applying Kruskal's multidimensional scaling procedure to the dissimilarity matrix $\bm{\Delta}$.\end{itemize}
\end{enumerate}
}
We used Metropolis-Hasting algorithms for updating the positions of items and the parameters,
and Gibbs sampling for updating the hyperparameters. Section B of the supplement provides additional details of the MCMC algorithm. 

\s\s

{\bf Remark.}
The postprocessing procedures (a) and (b) are motivated in Section 3.2.
In Step (a),
we follow the conventional approach to addressing the identifiability issues of latent space models by basing statistical inference on equivalence classes of positions,
using Procrustes matching \citep{Hoff:2002}.
In Step (b),
note that the matrix $\bm{\Delta}$ consists of dissimilarities rather than distances between schools,
because there is no guarantee that the dissimilarities of schools satisfy the triangle inequality.
That said,
Kruskal's procedure accepts dissimilarities as input and produces distances as output.
In other words,
the procedure described above produces an embedding of schools in $\mR^q$,
despite the fact that the matrix $\bm{\Delta}$ consists of dissimilarities rather than distances.

\subsection{Relations to existing approaches}

%\textcolor{black}{[mj:I cleaned and edited this section.]}

% An alternative approach is based on multivariate analysis of the scale-level scores, e.g., confirmatory factor analysis.
% However,
% the underlying assumptions of such approaches may be violated by the GEPS data: 
% e.g.,
% items may not measure a single (known) trait,
% %target trait (that is intended to measure);
% and not all correlations between the item response may be captured by the correlations between the traits measured with the scales (e.g., scale-level factor scores). 
% Recent studies have shown that test items are likely to be cross-loaded across scales \citep[e.g.,][]{Xu:18} and item responses are often conditionally dependent rather than independent \citep[e.g.,][]{Bolsinova:17}, suggesting that the assumptions of confirmatory factor analysis and other conventional approaches to analyzing educational assessment data may be too stringent.

A multilevel latent space model \citep{Sweet:13} has been proposed to handle school-level clustering for social network data, by including a random effect in a standard latent space model to capture variation across schools. In our proposed approach, we take one step further and  construct a school-level latent space, providing additional and more in-depth information on how schools are similar or different from each other in terms of their item or student network structures. 
Our multilevel network item response model simultaneously analyzes three types of networks, at the item, student, and school levels,  based on multilevel item response data. Thus, the proposed  model is a unique statistical tool that can identify three different types of sub-grouping (or clustering) structures that may be present at the item, student, and school levels.

% An alternative network modeling approach based on Ising graphical models \citep{RaWaLa10,Anetal12,BrKa20} have been proposed for item response data analysis \citep{vanBorkulo:2014, Kuris:2016, Epskamp:2017}. 
% That said,
% there are important differences between our network-based approach and the network-based approach based on Ising graphical models: 
% Ising-based methods estimates whether connections between items are absent or present and how strong these connections are,
% whereas we estimate distances between items as well as schools.

%While our approach,among other things, accomplishes tasks that can be accomplished by conventional mixture model approaches,
%our approach has advantages over conventional mixture model approaches.
%One may consider conventional latent variable modeling approaches to analyze the GEPS data. However, 
In addition, our approach has advantages over conventional latent variable modeling approaches. 
For example, 
multilevel exploratory item factor analysis (MEFA; \citealp{Muthen:94}) can identify clustering of items, i.e., factor structure, at a lower-level unit (within-group such as students) and at a higher-level unit (between-group such as schools); %based on within- and between-group data. 
%but MEFA cannot evaluate whether  the lower (e.g., students) and higher (e.g., schools) level units show the identical item clustering, i.e., factor structure; % (e.g., students and schools),
but MEFA cannot identify clustering of students and clustering of schools.  %clustering structures separately as in our multilevel network item response model. 
Multilevel mixture IRT models (MMIRT; \citealp{Vermunt:03}) can identify clustering of students and schools in terms of how the item parameters differ across clusters; however,  clustering of items may be unidentifiable at the same time.  
% In contrast, our multilevel network item response model detects clusters of schools based on how different schools are in terms of their item dependence (network) structures. Second, MMIRT focuses on identifying school-level clusters (latent classes) only, whereas our multilevel network item response model identifies clustering of all units of data, i.e., for items, students, and schools.  
% Another important difference of our approach from these mixture-modeling based approaches 
% is that with both MEFA and MMIRT the numbers of clusters is determined based on relative fit statistics comparisons, such as AIC and BIC, between several models (that differ only in the number of clusters or latent classes). In contrast, the multilevel network item response model does not require model comparisons as it automatically reveals item and student clustering structures once the model has been estimated from data. 
Alternatively, a multilevel item response theory (MIRT) model or a multilevel SEM model (MSEM) may be considered. % We mentioned in Section 2 that a number of assumptions of the MIRT model are likely violated with the GEPS data. 
Both MIRT and MSEM may offer information on how schools differ in terms of school-level average scores. % differences between schools. 
In contrast, our approach helps us examine how schools are different in terms of how individual students responded to individual test items.  
We demonstrated in Section 4.3 that an MIRT is unable to detect the differences among schools that our approach is able to detect with the GEPS data. 
%% as we demonstrate in Section 4.3.

% at the school-level latent trait or school effect information. Our model can do more, by allowing us to further examine differences between schools in terms of students responses to individual items. That is, different from MIRT or MSEM that  provide school-level averaged information only (averaged over all items or over items within sub-scales), our approach can detect school differences at the individual item level, which could be uncovered by MIRT or MSEM. 

%\clearpage

% A natural approach to analyze the GEPS data would be multilevel Rasch models \citep{fox:01, Kamata:01},
% because the GEPS data are multilevel item response data.
% That said,
% multilevel Rasch models are unable to detect the differences among schools that our approach is able to detect,
% as we demonstrate in Section 4.3.

\section{Analysis and results}
\label{sec:analysis}

\subsection{Method}\label{sec:method}
 
We apply the latent space model described in Section 3 to the GEPS data described in Section \ref{sec:description}. 
We assume that items have positions in $\mR^2$ and approximate the posterior by using the Markov chain Monte Carlo algorithm described in Section \ref{sec:posterior}. To detect possible non-convergence of the Markov chain Monte Carlo algorithm, we ran the Markov chain Monte Carlo algorithm five times, with starting values chosen at random. \textcolor{black}{Each Markov chain Monte Carlo run consisted of 50,000 iterations, with the first 25,000 iterations discarded as a burn-in. From the remaining 25,000 iterations, 2,500 samples were collected by retaining every tenth draw  from the posterior.} 
We set $\sigma_{\theta} = \sigma_{\mu} = 10$ and $a_{\tau} = b_{\tau} = .01$.  
Trace plots of draws from the posterior can be found in the supplementary materials (Section D). 
The differences between the five Markov chain Monte Carlo runs were negligible and the standard errors were less than $.05$ for all quantities. 
%as shown in the supplementary materials. 

\subsection{Model Fit}\label{sec:fit}

To evaluate model fit, we examined the predictive power of the model. The idea was to predict adjacency matrices $\bU$ based on the estimated model parameters---estimated by posterior means---and evaluate the predictive power of the estimated model in terms of sensitivity and specificity. It is worth noting that these predictive checks are not strictly posterior predictive checks in the Bayesian sense, but have the advantage of being less computation-intensive. 
These posterior predictions are in-sample predictions,
as is common in the literature on statistical network models,
including latent space models.

To evaluate predictive power of the proposed model, we defined sensitivity as TF/(TF + FN) and specificity as TN/(TN + FP), where TF: true negative, FN: false negative, TN: true negative, and FP: false positive. 
We also computed the overall accuracy as (TP + TN) / (TP + TN + FP + FN), where TP: true positive, TN: true negative, FP: false positive, and FN: false negative. 
We found that the sensitivity, specificity, and overall accuracy of the  proposed model turned out to be satisfying for $\bU$ with a sensitivity of 0.45 and a specificity of 0.82, and an overall accuracy was 0.710. 
We provide the sensitivity, specificity, and overall accuracy of individual schools in the supplementary materials (Section E). 
Taken together, the model fit assessment results suggest the predictive power, and thus, the fit of the estimated model was reasonable for the GEPS data. 

\subsection{Results}\label{sec:results}

We discuss results, with an emphasis on how the innovation and regular school systems were different in terms of item network and school network structures. We do not discuss student network structures here for the sake of parsimony; comparing within-school student  network structures across all schools is not   practical due to the large sample size. Plus, school-level network structures supply a sensible summary of student network structures across schools. 
%All other estimates, 
%such as the error variance for the linking function and student network structures, 
A summary of the person parameter estimates is provided in the supplement (Section C). 

\begin{figure}[htbp]
\centering
\begin{tabular}{c}
\includegraphics[width=0.7\textwidth]{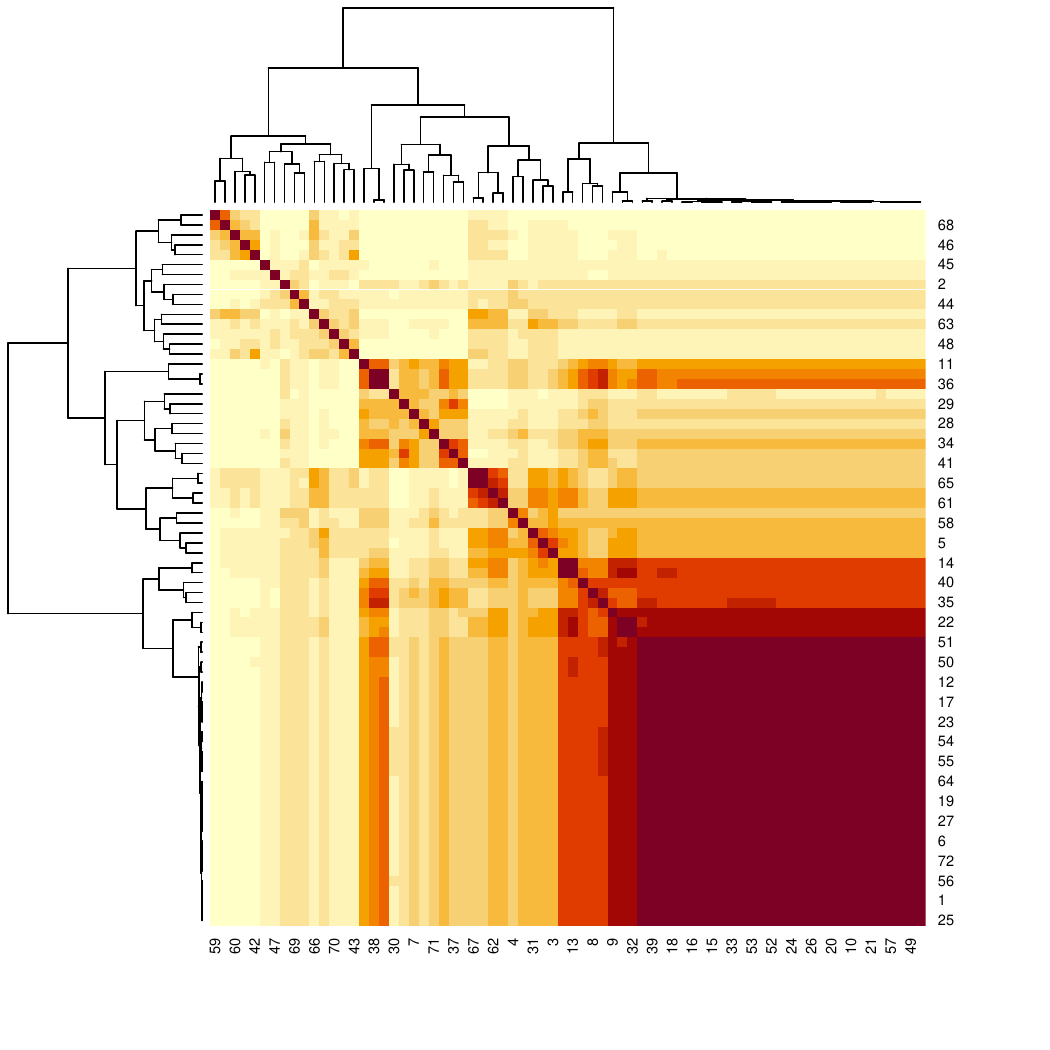}
\end{tabular}
\caption{\label{fig:ils_heat1}
Heatmap based on the estimated $\bm\mu = (\mu_{i,j})$ obtained from the single-group model. 
Numbers represent items. 
}
\end{figure}

\begin{figure}[htbp]
\centering
\begin{tabular}{c}
\includegraphics[width=0.5\textwidth]{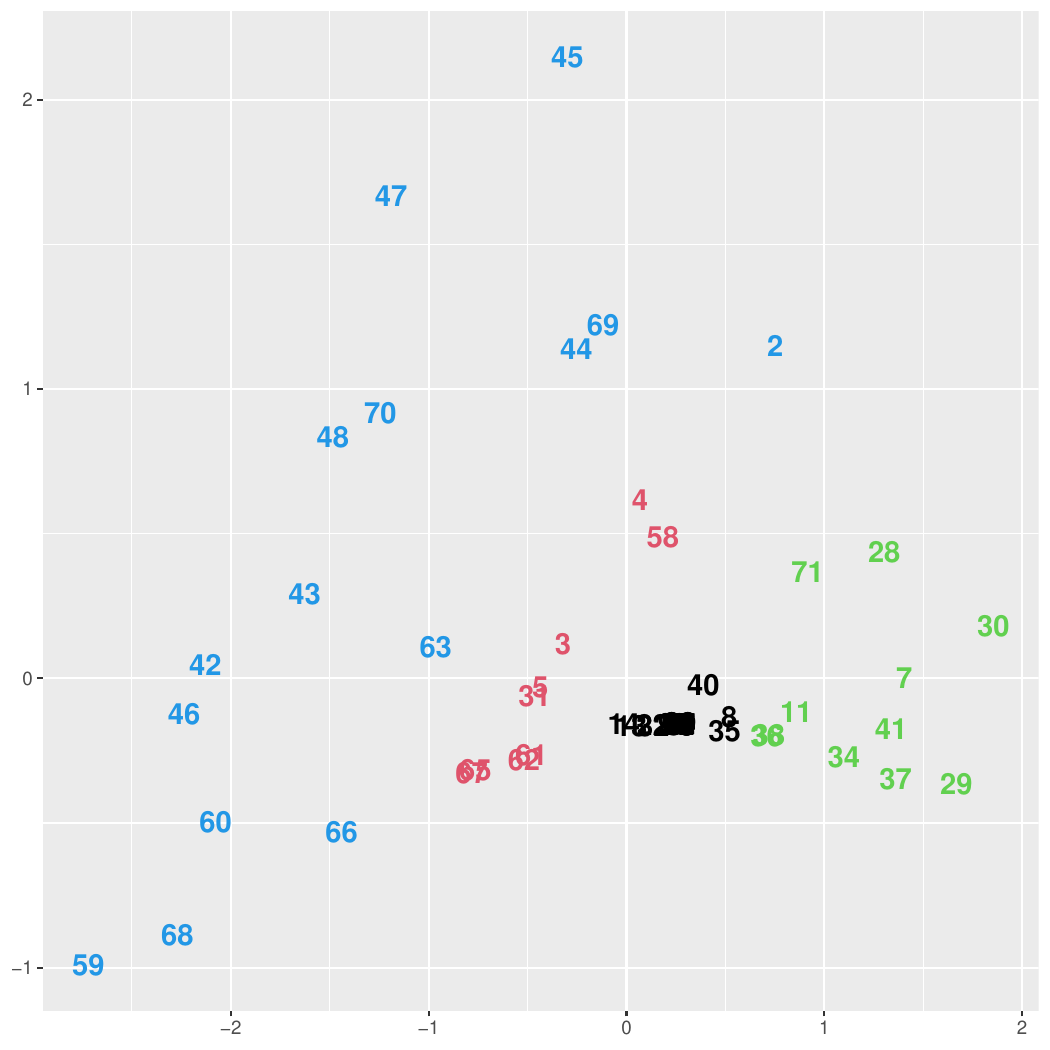}
\end{tabular}
\caption{\label{fig:ils_model1}
Embedding of items in $\mR^2$ based on the single-group model.
%, by applying Kruskal's multidimensional scaling \citep[MDS;][]{Cox:01} to $\bm\mu$.
Numbers represent items. 
Colors represent the clusters estimated from the dendrogram in Figure \ref{fig:ils_heat1}.
}
\end{figure}

\paragraph*{Item network structure}

%We examine the item network structure.
\textcolor{black}{Figure \ref{fig:ils_heat1}  shows a heatmap of $\bm\mu$ from the single-group model (see ``Item distances'' in Section 3.3).} The heatmap shows that a bulk of items is concentrated in a small region of the $\mR^2$ space. While most other items are not too far away from this item bulk, some are fairly distant from them. Those distant nodes, though, did not cause any computational problem  %, meaning that they were not isolated nodes in our analysis
%(All items had at least one ``agree'' response; 
(the minimum proportion of ``agree'' was 16.1\% for Item 59).

%\footnote{Note that these item groups are not based on their absolute positions in the latent space, because distances in $\mR^2$ are invariant to translation, reflection, and rotation of the positions.} 
%It is worth noting that such distant nodes in the item network could create problems, in that all non-isolated nodes may be concentrated in a small region of $\mR^2$, while the isolated nodes may be located at large distances from the bulk of the non-isolated items. We do not face such problems here,
%because there are no isolated nodes in the item network:

\begin{figure}[htbp]
\centering
\begin{tabular}{c}
(a) Regular schools \\
\includegraphics[width=0.6\textwidth]{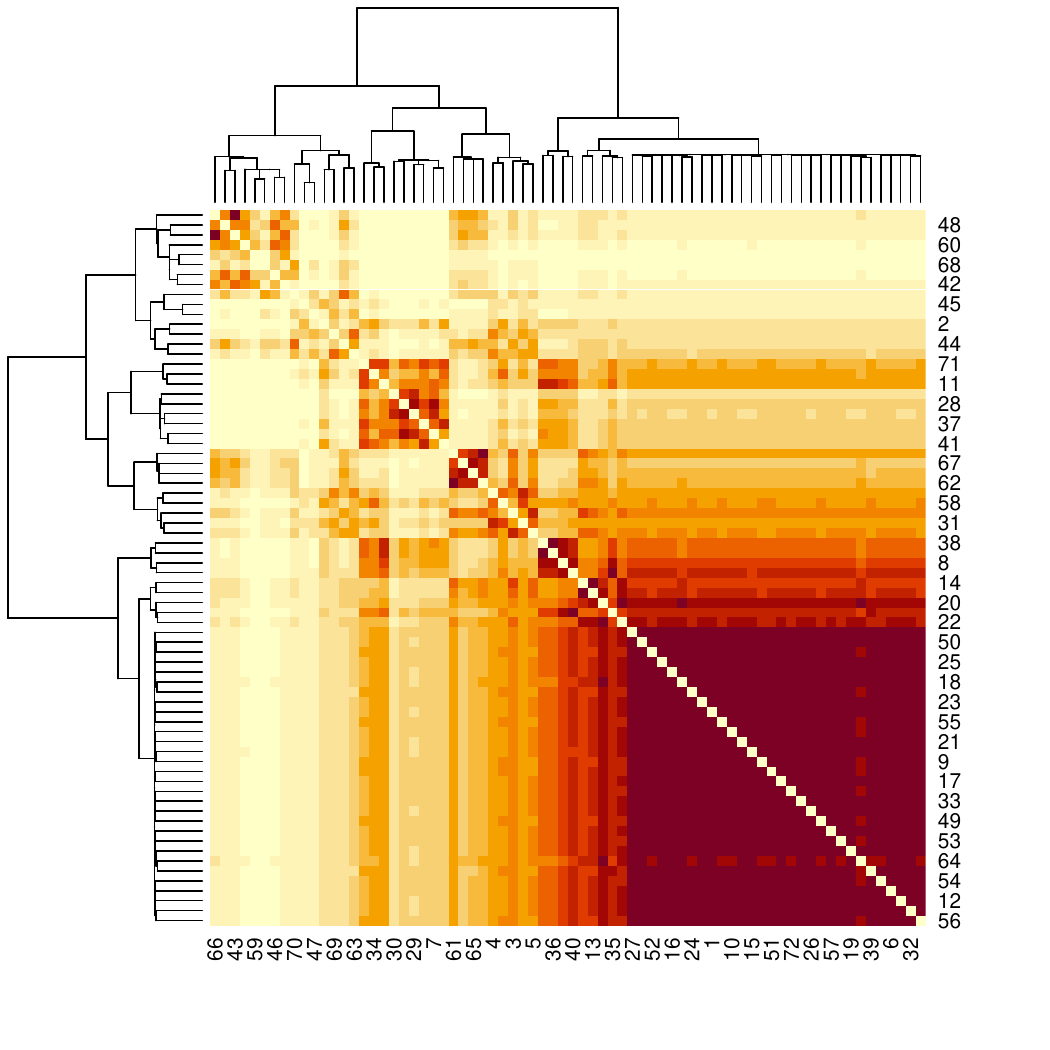} \\
(b) Innovation schools \\
\includegraphics[width=0.6\textwidth]{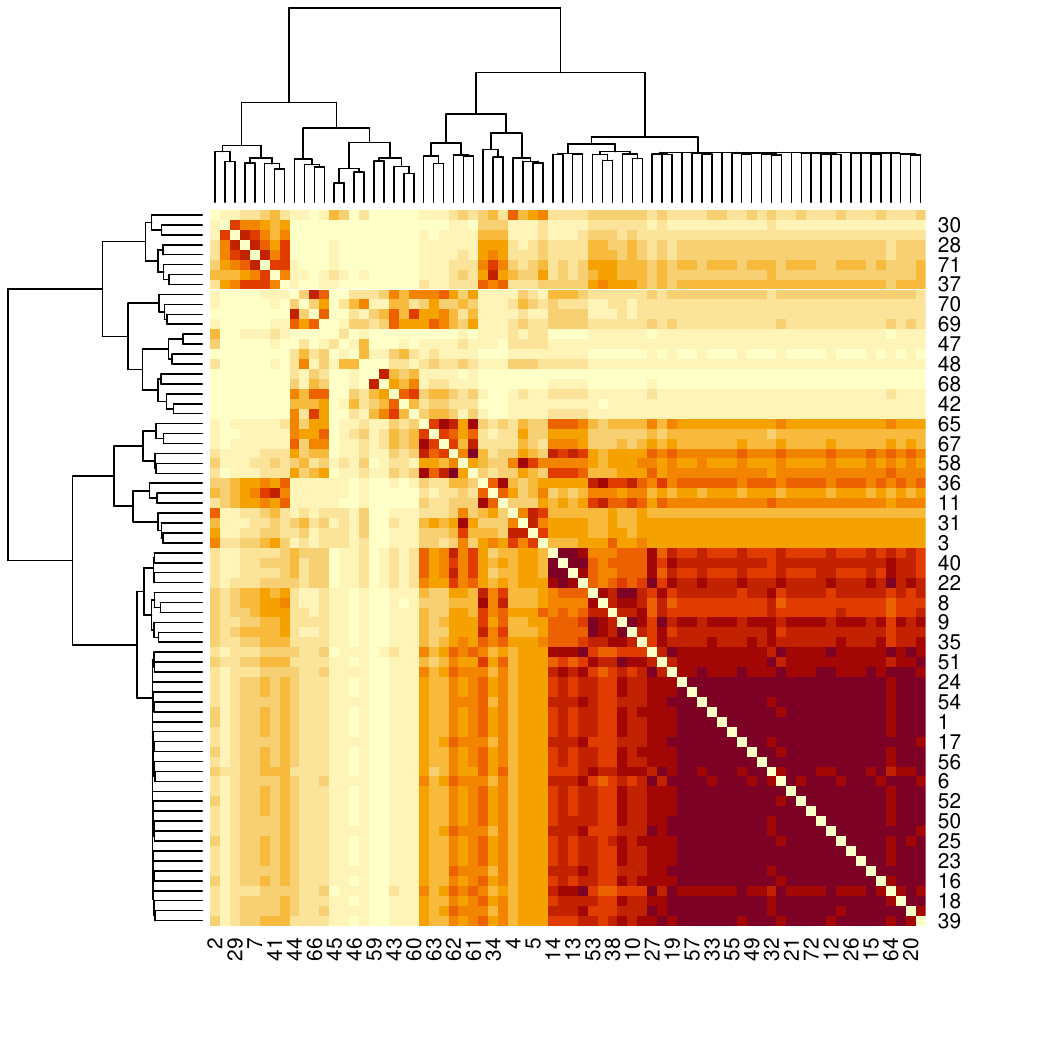} 
\end{tabular}
\caption{\label{fig:ils_heat2}
Heatmaps based on the estimated $\bm\mu = (\mu_{i,j})$ obtained from the multiple-group model: \break
(a) Regular schools and (b) Innovation schools. 
Numbers represent items.
}
\end{figure}

\begin{figure}[htbp]
\centering
\begin{tabular}{cc}
(a) Regular Schools  & (b) Innovation Schools \\
\includegraphics[width=0.45\textwidth]{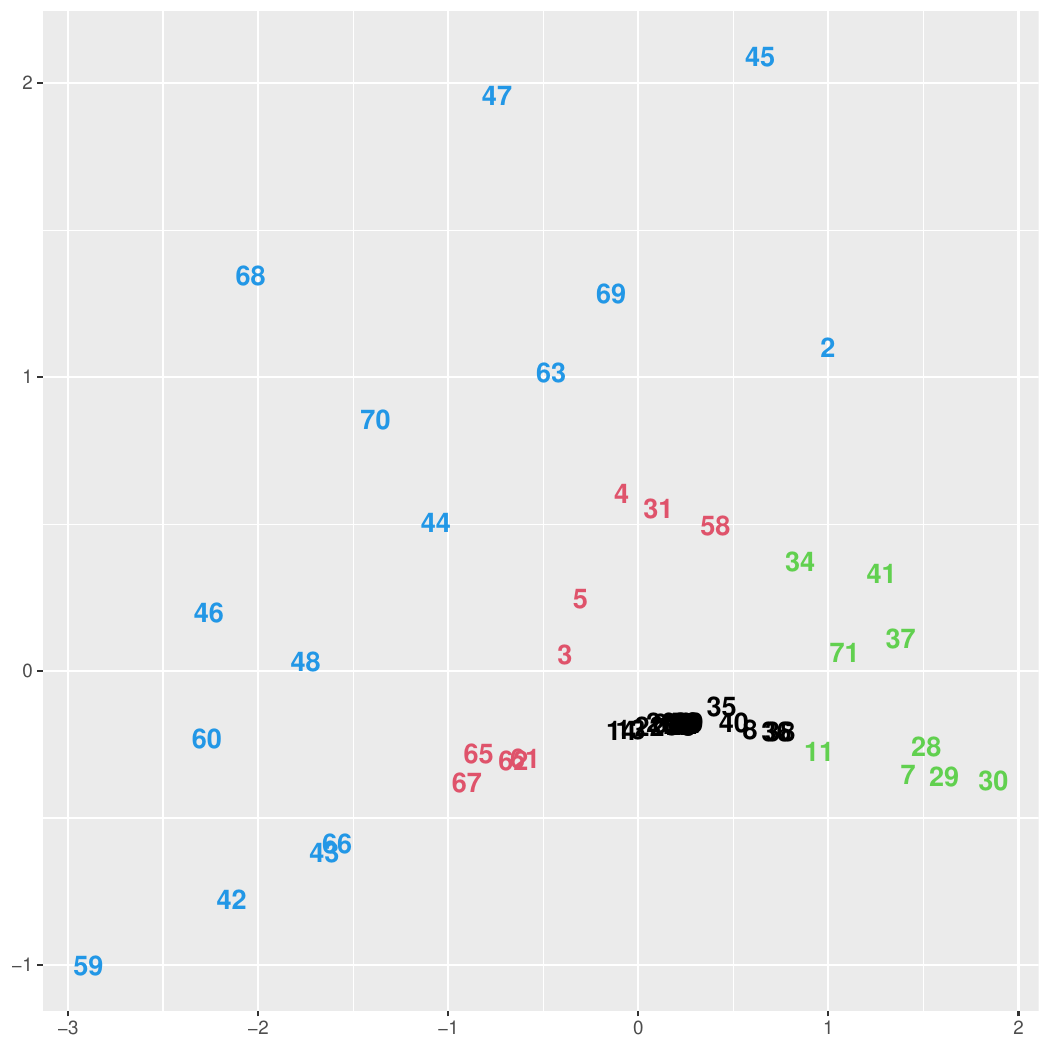} & 
\includegraphics[width=0.45\textwidth]{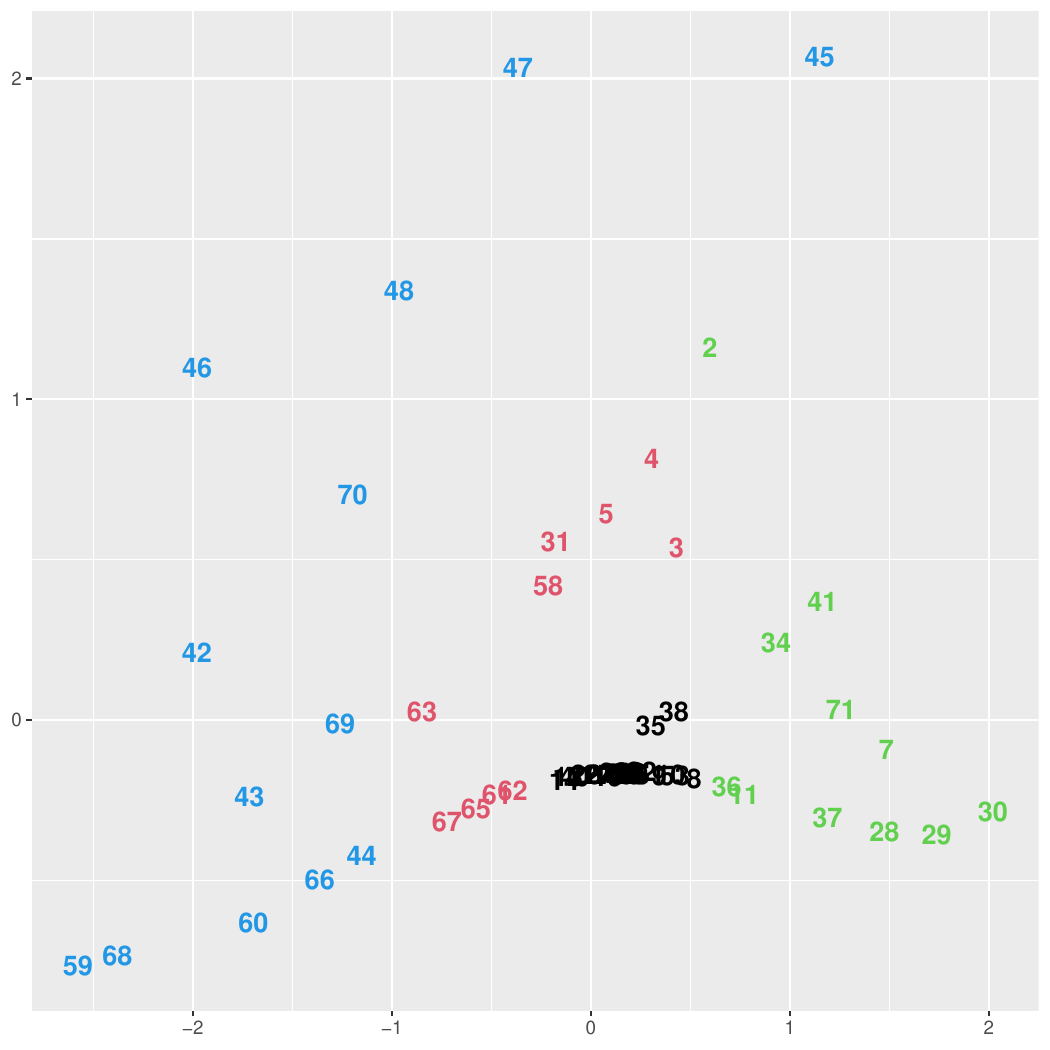}
\end{tabular}
\caption{\label{fig:ils_model2}
Embedding items in $\mR^2$ based on the multiple-group model:
(a) Regular schools and (b) Innovation schools. 
Numbers represent items. 
Colors represent the clusters estimated from the dendrograms in Figure \ref{fig:ils_heat2}.
}
\end{figure}

\textcolor{black}{
Figure \ref{fig:ils_model1} shows the latent space of the test items, constructed by applying MDS to $\bm\mu$.  
The numbers in black correspond to the item bulk discussed above with 
Figure \ref{fig:ils_heat1}; these items are located near the origin (0,0) of the $\mR^2$ space, meaning that a majority of  students in  the data positively endorsed the items in this group, which measure  ``Sense of Citizenship'',``Self-Efficacy'', ``Relationship with Friends'', and ``Self-Esteem''. %This suggests that  students follow the desirable direction in developing self-esteem and -efficacy, friendship, and civic consciousness. 
}
%\clearpage 
\textcolor{black}{
In the latent space,  we differentiated three item groups identified from the dendrogram (shown in Figure \ref{fig:ils_heat1}) in three different colors. %\textcolor{red}{[mj: could you clarify how the three  item groups were identified? did we apply spectral clustering?]} % located in the bottom right corner, the left of the center, and widely from the upper-middle to the bottom left corner. 
The items in green, placed in the east of the items in black, include several items measuring ``Self-efficacy'' and items measuring ``Sense of Citizenship ''  (I7),  ``Self-driven Learning (I37), and ``Self-understanding'' (I41).  The items in red,  located to the west of the items  in black, include items that measure ``Academic stress'' and ``Mental well-being''. The items in blue, spread in the far east of the black-numbered items, include items measuring ``Test stress'' and  ``Academic stress''. 
}

\textcolor{black}{
From the multiple  group model, we obtained two separate item network structures for innovation and regular school systems, which enables us to  investigate differences in the item network structure between the two school systems. 
Figures \ref{fig:ils_heat2} and \ref{fig:ils_model2} show the results. The two space configurations are nearly identical, except for two items --  Item 2 (``I am worried about everything.'') is in blue in the regular schools but in  green in the innovation schools, and Item 63 (``If you talk to your parents about my study, you get irritated rather than stable.'') is in blue in the regular schools but in red in the innovation schools. %, aside from the fact that the distances among schools in $\mR^2$ are invariant to translation, reflection, and rotation of the positions. 
The result suggests that overall, the response patterns of the students are nearly  indistinguishable between the two school types. 
}

% \textcolor{black}{
% Note that five items are assigned in different clusters between the innovation and regular school systems from the dendrograms in Figure \ref{fig:ils_heat2}. Those are Item 2 (``I am worried about everything.'', blue in the regular but green in the innovation) and Item 63 (``If you talk to your parents about my study, you get irritated rather than stable.'', blue in the regular but red in the innovation).
% }

\paragraph*{School network structure}

% \textcolor{black}{
% To examine the latent space of schools, we obtained the latent positions of individual schools by using $\Delta$ as explained in Section 3.3. 
% }

\begin{figure}[htbp]
\centering
(a)\\
\includegraphics[width=0.5\textwidth]{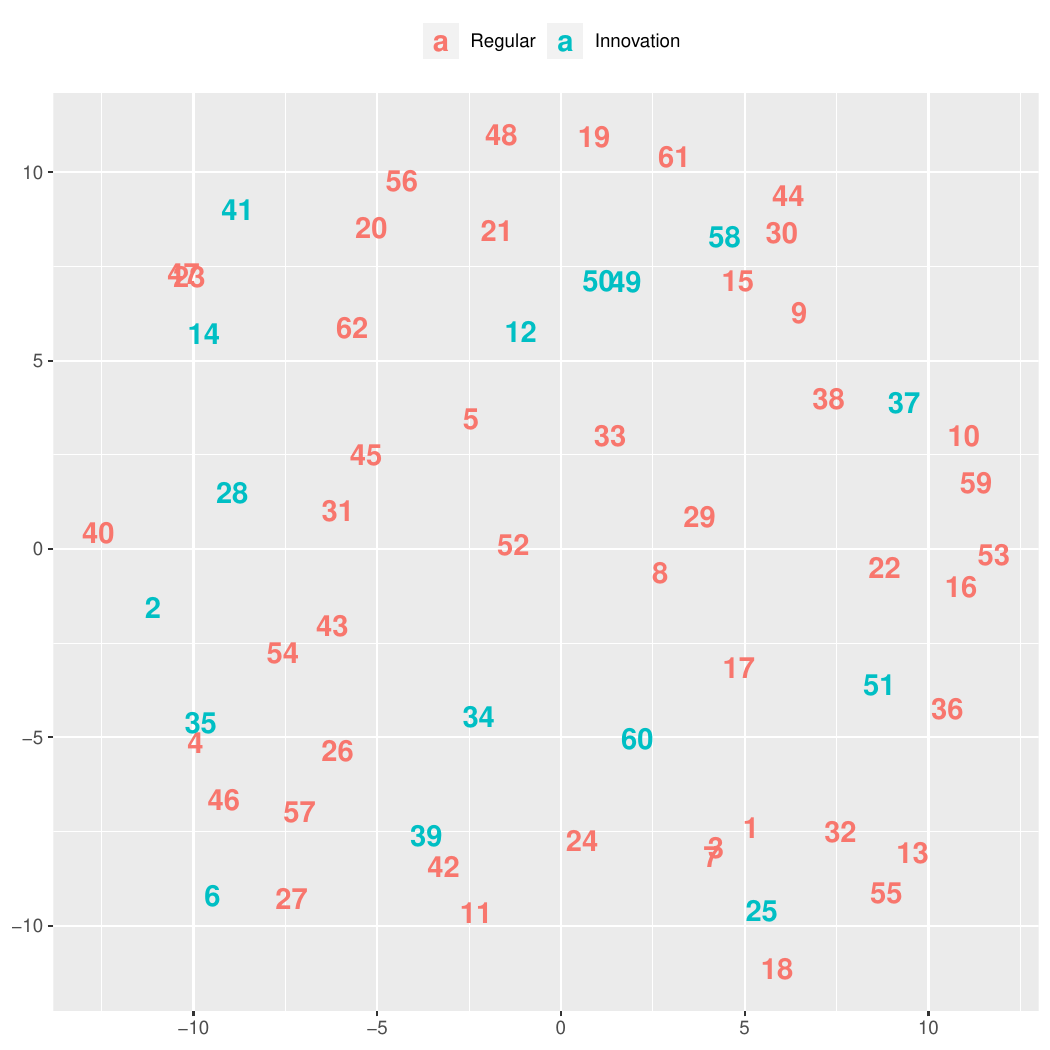} \\
(b)\\
\includegraphics[width=0.6\textwidth]{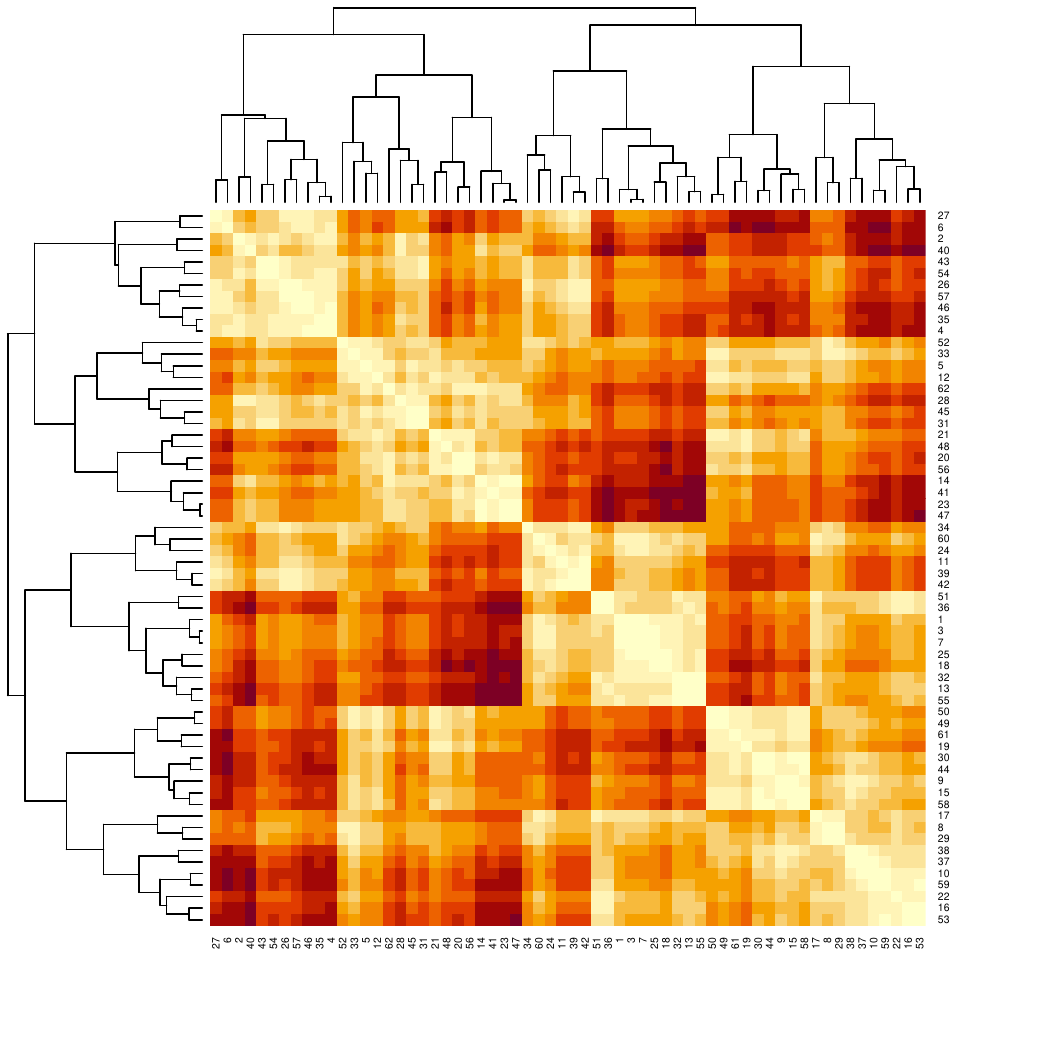}
\caption{\label{fig:school}
(a) Latent positions of individual schools constructed by the school-specific mean of the between-item distance measures $\Delta$. The procedure for embedding schools in $\mR^2$ is described in Section 3.3.
(b) A heatmap based on the estimated $\Delta$. }
\end{figure}

\textcolor{black}{
Figure \ref{fig:school}(a) shows %the heatmap and 
the latent space of the schools obtained based on $\Delta$, red and blue representing regular schools and innovation schools, respectively. 
Overall, the 62 schools are widely spared out  in the school latent space, but the locations of the innovation schools and the regular schools are rather mixed than separated from each other. 
The difference in the school latent positions reflect differences in the item dependence structures between schools. Thus, the results indicate that the two school systems are not clearly distinguishable in terms of the item dependence structures, suggesting % (i.e., there are no clear patterns in the school locations based on the school types).  
that the school type (innovation vs. regular) might not be  a main deriving factor that explains school differences  in the relationships between the test items (which is the result of how students responded to the test items). %item dependence structure.
}

\textcolor{black}{
%In Figure \ref{fig:school}(b), only a few schools are located near the origin and all others are formed multiple clusters. 
The dendrogram in Figure \ref{fig:school}(b) shows four school clusters: a upper-left group, a upper-right group, a bottom-left group, and a bottom-right group. For close inspection, we picked one school from each cluster: Schools 1, 2, 56, and 10, respectively. Schools 10 and 2 are far away from each other, located at the north-east side and south-west side of the latent space. 
Schools 56 an 1 are also far apart from each other, located at the north-west side and south-east side of the space. 
School 2 is the only innovation school among  the four selected schools.  }

\begin{figure}[htbp]
\centering
\begin{tabular}{cc}
(a) School 10  & (b) School 1  \\
\includegraphics[page=10,width=0.45\textwidth]{Figure/item_ls.pdf} & 
\includegraphics[page= 1,width=0.45\textwidth]{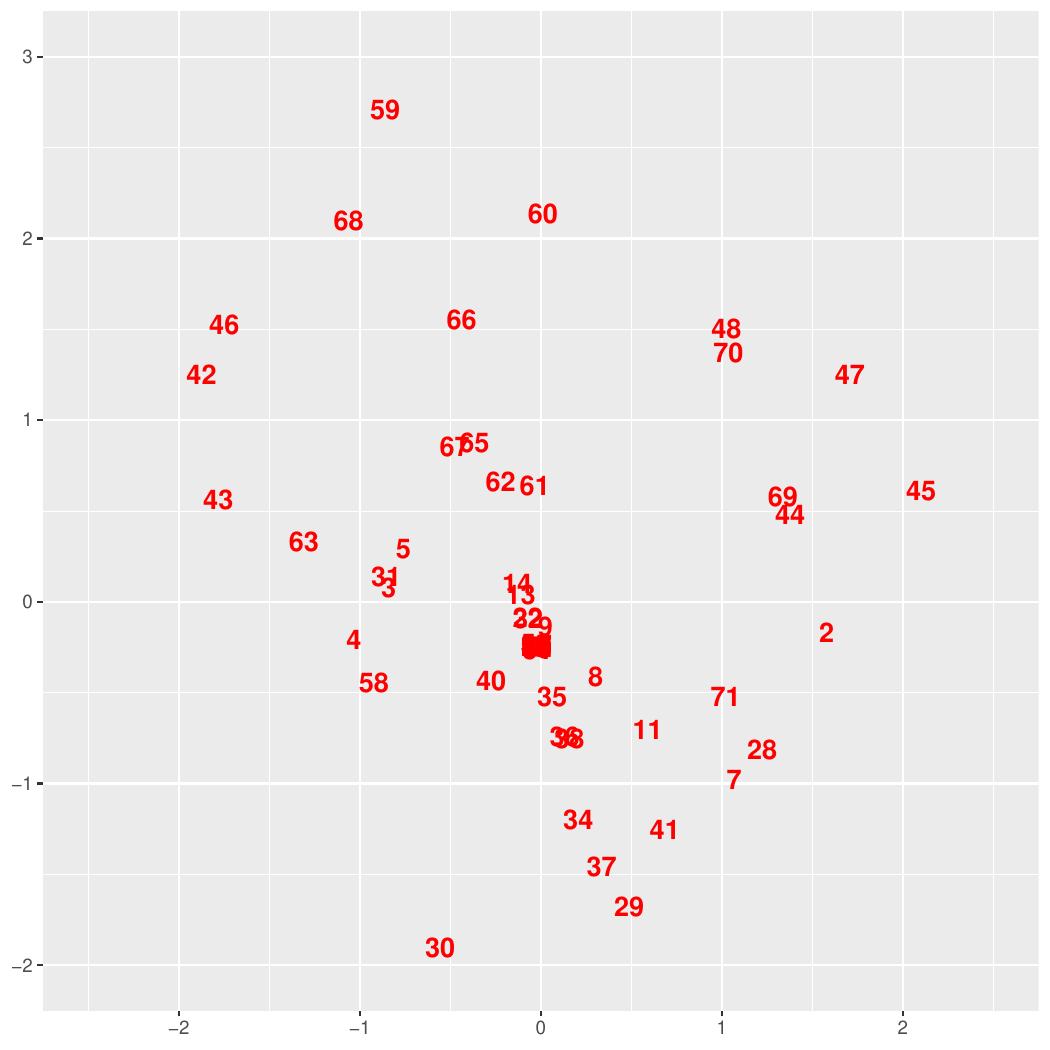} \\ 
(c) School 2 & (d) School 56 \\
\includegraphics[page= 2,width=0.45\textwidth]{Figure/item_ls.pdf} & 
\includegraphics[page=56,width=0.45\textwidth]{Figure/item_ls.pdf}
\end{tabular}
\caption{\label{fig:sch2}
Student and item latent spaces for Schools 1, 2, 10, and 56. Red numbers represent item latent positions, respectively.}
\end{figure}

Figure \ref{fig:sch2} presents the item latent spaces of the four selected schools. %\textcolor{red}{[mj: please check if the four schools are 1,2,56, and 10 in this figure. The figure labels and captions say different school  numbers]} 
The item latent spaces of all other schools are provided in the supplementary materials (Section F). It is interesting that a virtual line may be drawn from Item 59 to Item 29 (from top left to bottom  right) in all four latent spaces; and items near the virtual line are nearly identical in the four latent spaces. However, the positions of most other items outside this line appear  different in the four latent spaces. 
For example, items on the east to the origin (e.g., Items 2, 3, 11, 30, 42, 44, 45, 48, 69, 70, 71) are positioned differently in Schools 2 and 10, meaning that the relationship between the items, in other words, how students  conceived and responded to  those items were different in the two schools. 

For a better understanding of the school-specific latent spaces examined above, we checked the proportion of ``agree'' (positive responses) of the two schools, School 2 and 10 in Figure \ref{fig:sch2}. %The result is shown in Figure \ref{fig:sch2}. % for the result. 
%\textcolor{red}{[mj: 1) did we pick school 2 and 10? the figure label and caption say different things. 2) also, can we show only (c)? (c) seems more clear than (a) and (b),  3) also, can we explain (and please remind me) why we picked only 2 schools here? we picked four schools before this. which one is renovation and which one is regular school? ]}
%The item dependence structures shown in Figure \ref{fig:sch2} 
As expected, the figure shows that there are notable differences between the two schools in terms of how students responded to the test items. 
%students in School 10 tend to ``agree'' as frequently as ``disagree'' with items, whereas 
% there  are notable discrepancies in the ``agree'' proportions between the two schools students.
Overall, the proportion of positive responses appears higher in School 2 than School 10 for most items, while for some item, such as Items 2--6, 28--30, 45, 70, the proportion of ``agree'' is higher in School 10 than School 2. 
%This is likewise supported by the proportions of agreements shown in Figure \ref{fig:prop}(a) and (b). We take a closer look at students' response patterns in \ref{fig:prop}(c) that displays the proportions of ``agree''  across 72 items.  
%School 10 shows a different response pattern compared to School 1. For instance, 
%Specifically, there  are notable discrepancies in the ``agree'' proportions between the two schools students.
%in School 10 appear to select ``agree'' far  less often than  School  2 for items 8--27, 41--49, and 50--70.  
Those items measure sense of citizenship, self-efficacy and reverse-coded academic stress, and they roughly correspond to the items that are positioned differently in the school-specific latent spaces shown in Figure  \ref{fig:sch2}. 
This inspection shows that schools located in different regions of the school-specific latent space are indeed different in terms of their students' response patterns. 

All in all, our analysis results  tell us that there are substantial differences across schools in terms of how the students conceived and responded to items measuring their mental well-being, but  the differences were not necessarily driven by whether or not  the schools adopted the innovation school system. 
%In addition, students in School 10 also select ``agree'' for items 29--34, items 44--46, items 55--56 more frequently than students  in School 1, while those items are about high self confidence and high level of test-related stress. 
%This suggests that students in School 10 might appreciate citizenship and relationships with others less than students in School 2. 
%\textcolor{red}{[mj: also, how can we relate this result to what the latent space shows in figure  6?  The connection seems missing ]} 
%In addition, students in School 10 may have more self-confidence but---at the same time---seem to experience higher levels of test-related stress than students elsewhere. 

% \vspace*{-.25cm}
% \vspace*{.25cm}
% \textcolor{black}{
% The above results demonstrate that %the school network structure identified by our approach 
% the proposed approach helps us identify differences between schools in terms of the response patterns of their students. In our analysis, whether the schools adopted the innovation school program or not did not appear to derive substantial differences between schools. 
% %, although we did detect ``outliers" of interest, to mention School 10. 
% }

\begin{figure}[htbp]
\centering
\includegraphics[width=0.7\textwidth]{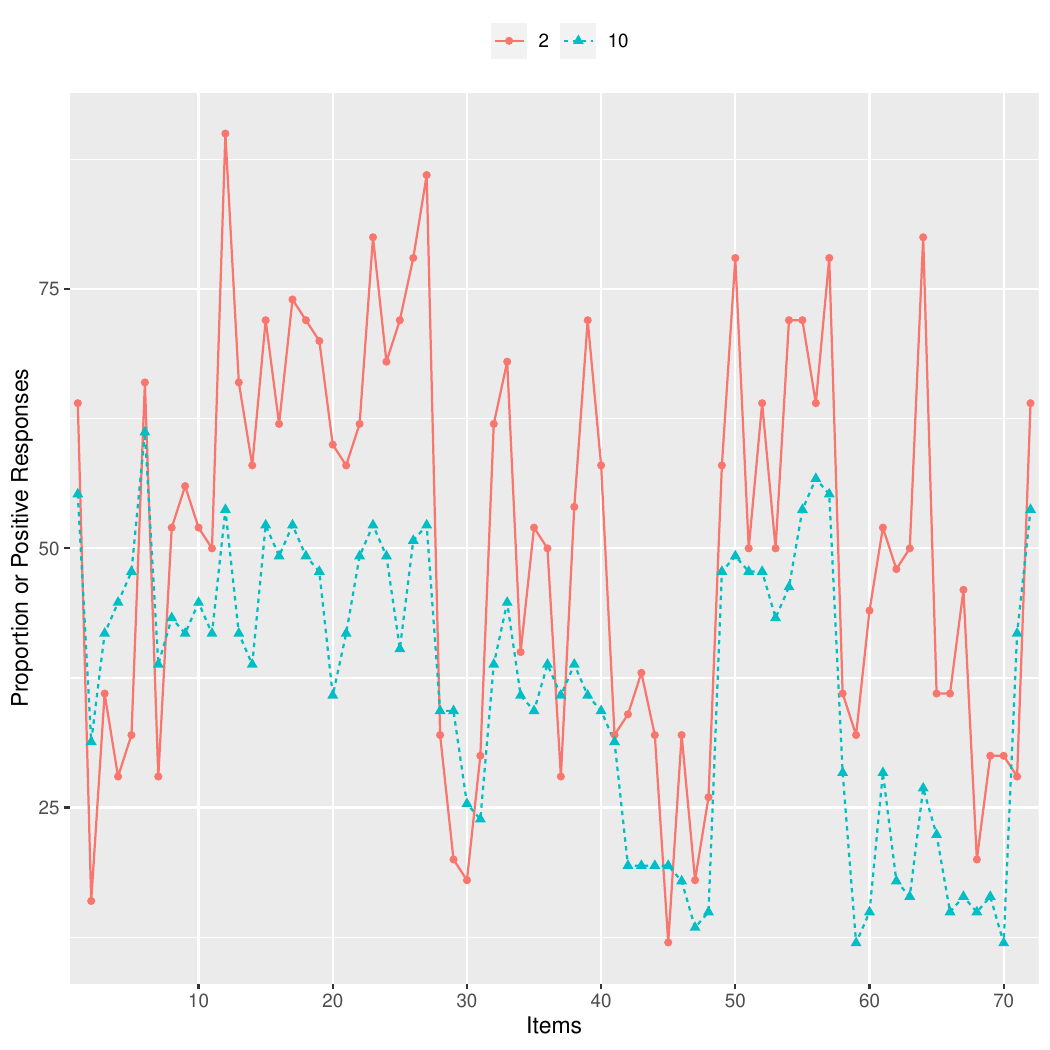} 

% \begin{tabular}{cc}
% (a) School 2  & (b) School 10  \\
% \includegraphics[page= 2,width=0.45\textwidth]{Figure/prop_agree.pdf} &
% \includegraphics[page=10,width=0.45\textwidth]{Figure/prop_agree.pdf} \\ 
% \end{tabular}
% \begin{tabular}{c}    
% (c)  \\
% \includegraphics[width=0.7\textwidth]{Figure/prop_positive_each.pdf} 
% \end{tabular}
\caption{%(a) and (b) Histogram of the proportion of  ``agree'' for School 10 and School 28. 
Proportion of ``agree'' across 72 items in Schools 2 and 10. }
\label{fig:prop}
\end{figure}

\paragraph*{Comparison: multilevel Rasch model}

We applied a multilevel Rasch model to the GEPS data, with the school program as a school-level covariate, to compare our approach to a conventional approach. We chose  a multilevel Rasch model for a comparison because this model assumes the same latent variable structure (with a single latent variable) at the student level as the estimated model, and allows us to make inferences about differences in student responses between the two school types of interest. 
A multilevel Rasch model posits that the log odds of the probability of a positive response by student $k$ in school $m$ to item $i$ is
\beno
    \mbox{logit} (\mathbb{P} (X_{m,i,k} = 1 \mid \beta_i,\, \gamma,\, \delta_m,\, \theta_{m,k}) 
    &=& \beta_i + \gamma\; I_m + \delta_m +  \theta_{m,k},
    %\;  \theta_m \sim N(0, \tau_{sch}^2), \;  \theta_{mk} \sim N(0, \tau_{stu}^2), 
\ee
where $\beta_i \in \mR$ and $\delta_m \in \mR$ are the item-specific and school-specific intercepts and $\theta_{m,k} \in \mR$ are the student- and school-item-specific effects, respectively. The parameter $\gamma \in \mR$ represents the deviation of the Innovation schools' effects from the regular schools' effects (where $I_m$ represents the binary indicator for innovation schools).

\textcolor{black}{The multilevel Rasch model specified above was estimated by Bayesian Markov chain Monte Carlo methods, using R package MCMCglmm \citep{Hadfield:10} with 13,000 iterations, 3,000 burn-in iterations, and recording every $10$-th draw from the posterior. There is little evidence that the two school programs differ in terms of the overall school means, 
because the 95\% posterior credible interval is [-0.139,  0.301] (which includes $0$). 
The posterior means of school-specific intercept $\delta_m$ ranged from -1.27 to 0.82 across 62 schools with the mean of 0.01. %The school-specific intercepts of Schools 1, 2,  56,  and 10 do not stand out:
The posterior means and 95\% posterior credible intervals of the school-specific intercepts are 
%-0.21 [-0.36, -0.07], 
-0.25 [-0.46, -0.01], 
%0.06 [-0.08, 0.22], 
and  -1.27 [-1.42, -1.11] for School 2 and 10, respectively. That is, this result tells us that  students  in School 10 on average showed a lower level of mental well-being than students in School 10, without giving in-depth information about how the schools differ in terms of student responses to the  well-being questions. 
%This result demonstrates that conventional approaches focusing on differences between school means -- such  as conventional multilevel models -- is unable to show differences between in depth in terms of students'  response patterns. % with our approach.
}

\section{Conclusion}
\label{sec:conc}

The Korean school innovation program, implemented in the last 9 years in Korea, has been under public scrutiny due to widespread skepticism about its effectiveness. 
Most research on the school innovation program has relied on simple comparison methods and produced mixed results about the programs' effects on students' non-cognitive outcomes, calling for a more comprehensive study that can shed new light on the issue. 

Motivated by this need, we proposed a novel analytic approach that allows us to explore differences between the innovation and regular school programs from a fresh angle in terms of their students' non-cognitive outcomes.  
The innovation of the  proposed multilevel network modeling approach lies in that 
a latent space modeling technique, originally developed for network data, was adapted to analyze multilevel item response data for the purpose of examining networks (similarities and dissimilarities) of items, students, and schools, in addition to the item and student parameters.  % could be examined. 
This method has not only technical advantages compared with traditional multilevel modeling methods, but also provides substantive merits by allowing us to investigate more subtle differences between innovation schools and regular schools than other methods cannot  offer. %that has never been taken. 

Upon an inspection on the item network structure of the GEPS data, we did not find strong evidence that   innovation school students were distinguishable  from regular school students in terms of the  item network structure. That is, the students in the innovation school system were not different from the students in the regular school system in terms of how they conceived and responded to individual test items measuring  positive self-image or mental well-being. 
%meaning that the two school systems were not different in terms of how all the test items were conceived and responded by their students. 

% showed higher levels of desirable attributes that are reflected in some items, such as a sense of democracy, citizenship, autonomy (that the program intended to foster) compared with  regular school  students. 
% An interesting finding is that 
%  innovation school students seem to believe that honest and truthful communications as well as concerns about tests are related to relationships with other people, whereas  regular schools students believe that those values and concerns are connected with themselves rather than with others. 

% In terms of the item network structure, however, innovation schools were found to be indistinguishable from regular schools, meaning that the two school systems were similar in terms of how all the test items were conceived and responded by their students. 
% We identified a common phenomenon that  Korean high-school students, both in innovation and regular high schools, were under enormous stress and pressure about their academic performance. 

Even so, our approach did reveal that some schools were  rather different from the others in terms of the item network structure, although the differences do not appear to stem from whether or not the schools adopted the innovation program. 
Those differences were not detected by conventional approaches, such as multilevel models, as shown in this paper. 

It would be beneficial to apply a formal test to make a selection between the proposed model and simpler, multilevel models. However, due to the substantial differences in the basis and structure of the two kinds of models, developing a formal comparison test is a challenging task and it  is beyond the scope of the current paper. 
Simpler alternatives such as multilevel models  may have computational  advantages over our approach. However, the proposed approach provides a unique opportunity to examine the  differences between the two school types in the current  context  from a unique angle, in terms of how students responded differently to specific test items. 
For instance, suppose two schools were similar in terms of average student performance in a mathematics test; but in one school students performed better with calculation items than data analysis  items, while in the other school,  students performed better with data analysis items than calculation items. 
Therefore, despite  potential computational cost, the proposed approach can be beneficial for the  opportunity that enables us to make an in-depth investigation and comparisons among schools. 

\subsection*{Source code and data}

The {\tt R} source code used in Section 4 can be found on GitHub:
\begin{verbatim}
https://github.com/Jonghyun-Yun/HiNIRM
\end{verbatim}
The data cannot be shared without the consent of the government of South Korea.
Interested readers are referred to the English-language website of the Gyeonggi Institute of Education in South Korea:
\begin{verbatim}
https://www.gie.re.kr/eng/content/C0012-04.do
\end{verbatim}

\subsection*{Acknowledgements}

We are grateful to two anonymous referees, the Associate Editor and the Editor for constructive comments and suggestions that have led to substantial improvements of the manuscript.
Ick Hoon Jin was partially supported by the Yonsei University Research Fund of 2019-22-0210 and by Basic Science Research Program through the National Research Foundation of Korea (NRF 2020R1A2C1A01009881). Michael Schweinberger was partially supported by NSF awards DMS-1513644 and DMS-1812119 and ARO award W911NF-21-1-0237 (75549-NS). Lizhen Lin would like to acknowledge the general support from NSF grants IIS-1663870 and DMS-1654579 and Darpa grant N66001-17-1-4041.

\bibliographystyle{Chicago}
\bibliography{reference}

\end{document}